\documentclass{ws-ijmpa}

\begin{document}


\markboth{} {}
%
\catchline{}{}{}{}{}
%

\title{SOME PROPERTIES OF THE CENTRAL $\pi^-$-MESON CARBON INTERACTIONS AT 40 GeV/C}

\author{\footnotesize M.K. Suleymanov$^{\star,+}$$^,$\footnote{E-mail address:mais\_suleymanov@comsats.edu.pk
}~, E.U. Khan$^{+}$,K.Ahmed$^{+}$, Mahnaz Q. Haseeb$^{+}$, Farida
Tahir$^{+}$, Y.H. Huseynaliyev$^{+}$, M. Ajaz$^{+}$,
K.H.Khan$^{+}$,Z.Wazir$^{+}$ }

\address{
$^{\star}$Joint Institute for Nuclear Research, Dubna, Moscow region 141980, Russia\\
$^{+}$Department of Physics COMSATS Institute of Information
Technology,Islamabad, 44000, Pakistan\\}

\maketitle

\begin{history}
\received{Day Month Year}
\revised{Day Month Year}
\end{history}

\begin{abstract}
We discuss some properties of the central $\pi^-$-meson carbon
reactions at 40 GeV/c. While these results were obtained many years
ago they have not been explained completely.  We attempt to
interpret following: results regime change on the behavior of some
characteristics of the events as a function of the centrality;
anomaly peak on the angular distributions of the slow protons
emitted in these reactions; charge asymmetry on the $\pi^-$-mesons
production in the back hemisphere in lcs.

Understanding of the results could help to explain the new ones
coming from the modern central experiments at high and
ultrarelativistic energies.

\keywords{central; pi-meson; carbon.}
\end{abstract}

\ccode{PACS numbers: 11.25.Hf, 123.1K}

\section{Introduction}

When we pass from hadron-light nuclei interactions to
nuclear-nuclear interaction at relativistic and ultrarelativistic
energies we get the new possibility: to create   the high density
and high temperature  hadronic matter and get the information on
properties of matter under extreme conditions.  In such new
conditions the volume of information increases sharply as well as
the background information. The latter can grow faster than useful
signal information due to the reason that the number of secondary
multiparticle interactions become more and more (it is very
essential in case of central collisions). Simultaneously we can lose
some information and may miss some changes that could be important
to understand the dynamics of the interaction. So it is important to
know what kind of information was lost or changed and how we can
restore it. Best way to do it would be to turn to hadron-nuclear
interactions and to try to use the connections between the
properties of the hadron-nuclear interactions and the ones of the
relativistic and ultrarelativistic heavy ion collisions. In this
case the new ideas coming from the latter can help us to understand
hadron-nuclear interactions too.

The goals of the paper are:to show some properties of the central
$\pi+A$-reactions and the connections of ones with the properties of
the relativistic and ultrarelativistic nucleus-nucleus collisions;
to explain qualitatively the results on:regime change on the
behavior of some characteristics of the events as a function of the
centrality; anomaly peak on the angular distributions of the slow
protons emitted in these reactions; charge asymmetry on the
$\pi^-$-mesons production in the back hemisphere in lcs, using new
ideas coming from the relativistic and ultrarelativistic
nucleus-nucleus physics.

\section{Regime change in central experiments}

The central  experiments demonstrate the point of regime change and
saturation in the behavior of some characteristics of the events.
This phenomenon could be connected with fundamental properties of
the strongly interacting matter.

Here we would like to discuss some results connected with the
properties of the central $\pi^{-12}C$-interactions at 40 GeV/c. The
data has been obtained with the 2-m propane bubble chamber of LHE,
JINR~\cite{1}. To fix the centrality of collisions the number of
identified protons ( $N_p$ ) were used. Fig.~\ref{fj1} shows a
number of the $\pi^{-12}C$-interactions ($N_{star}$ ) as a function
of the $N_p$~\cite{2}. One can see the regime change in the behavior
of the values of  the $N_{star}$  as a  function of the $N_p$ near
the value of the $N_p=4$. This could be used to select the
$\pi^{-12}C$-reactions with total disintegration of nuclei (or
central collisions).

In paper~\cite{3} the results from BNL experiment E910 on pion
production and stopping in proton-Be, Cu, and Au collisions as a
function of centrality at a beam momentum of 18 GeV/c are presented
(see Fig.~\ref{fj2}). The centrality of the collisions is
characterized using the measured number of "grey" tracks,
$N_{grey}$, and a derived quantity $\nu$  , the number of inelastic
nucleon-nucleon scatterings suffered  by the projectile during the
collision. In Fig.~\ref{fj2} is plotted the values of average
multiplicity for $\pi^-$ - mesons ( $<\pi^--Multiplicity>$) as a
function of $N_{grey}$ and $\nu$ for the three different targets. We
observe that $<\pi^- - Multiplicity>$ increases approximately
proportionally to centrality and for all three targets at small
values of $N_{grey}$ and $\nu$ then they saturate with increasing
centrality in the region of more high values of $N_{grey}$ and $\nu$
. Solid line in figure shows the expectations for the $<\pi^--
Multiplicity>$ ($\nu$ ) based on the wounded-nucleon (WN)
model~\cite{3} and with dashed lines, does a much better job of
describing $p-Be$ yields than the WN model.

The results of the Ref.~\cite{5}  have shown the regime changes on
the behavior of the event numbers for the $dC$, $HeC$ and
$CC$-interaction at 4.2 A GeV/c~\cite{5} as a function of the
centrality.

It is very important that the regime change has been indicated even
on the behavior of heavy flavor particles production in
ultrarelativistic heavy ion collisions as a function of
centrality~\cite{6}.

A mechanism to explain the phenomena could be the percolation
cluster formation which is discussed to explain the results coming
from the ultrarelativistic heavy ion collisions~\cite{percolj}.
Percolation clusters may be formed in these interactions independent
of the colliding energy but the structure, maximum density and
temperature of hadronic matter may depend on colliding energy and
masses in the cluster framework.

\section{The angular distributions of the protons emitted in
central collisions}

In Fig.~\ref{fj6}. the angular distribution of protons with momentum
less than $1.0 GeV/c$ is shown for $\pi^{-12}C$-reactions at 40
GeV/c with total disintegration of carbon  nuclei (central
collisions)~\cite{7}. We can see some peak at angles close to
$60^0$. This result was confirmed by the results obtained at
investigation of the angular distributions of ones emitted in
$\pi^{-12}C$-interaction (at $5 GeV/c$)~\cite{8}.

In the paper~\cite{9} was presented the angular distribution for
slow protons emitted in the central $He+Em$-( at 2.1 A GeV), $O+Em$
- (2.1 A GeV) and $Ar+Em$ - (1.8 A GeV) collisions . Some wide
structure was observed in these distributions.

So one can say that the angular distributions of slow particles
emitted in central $\pi$-meson and light nuclei interactions with
nuclear targets indicate a structure - peak at some angle. We think
that the peak could be result of the formation and decay of the
percolation cluster.

\section{Charge asymmetry on the $\pi$-meson production in the back hemisphere in
lcs.}

 In Ref.~\cite{13} charge asymmetry in $\pi^{-12}C$-reaction at 40 GeV/c  was observed in
$\pi^{-12}C$-reactions at 40 GeV/c: number of positively charged
$\pi$-mesons were greater than negatively charged ones for back
hemisphere in lcs and it becomes more  with momentum of the
particles. It was obtained that these particles produced mainly in
central events (or events with total disintegration of nuclei).
Recently some papers appeared with very interesting idea connected
with charge asymmetry. In the Ref.~\cite{14} authors tried to show
that in the presence of a magnetic field QCD predicts topological
charge changing transitions can separate quarks according to their
electric charge along the direction of the magnetic field. This is
the so-called Chiral Magnetic Effect. They argued that  it might be
possible to observe the Chiral Magnetic Effect in heavy ion
collisions. May be it is fantastic idea to support that the
observation of charge asymmetry for  pi-meson production in back
hemisphere in lcs  in $\pi^{-12}C$-reactions at 40 GeV/c connected
with the difference of charges of  u and d quarks. But if we turn to
percolation cluster idea we can say it may be due to  the
percolation cluster strongly charged system and its motion could
lead to appearance of a magnetic field which could be a reason of
charge asymmetry for the $\pi$-mesons production in the back
hemisphere in lcs.

{\bf Captures of figures}

Fig.1 A number of the $\pi^{-12}C$-interactions as a function of the
$N_p$-centrality.

Fig.2  The average multiplicity of the $\pi^-$-mesons.

Fig.3 The angular distributions of protons emitted in
$\pi^{-12}C$-interaction (at 40 GeV/c) with total disintegration of
nuclei.

Fig.4  Momentum distributions of the $\pi^\pm$-mesons produced in
$\pi^{-12}C$-interaction at 40 GeV/c.


\begin{thebibliography}{0}

\bibitem{1} BBCDHSSTTU-BW Collab. (A. U. Abdurakhimov {\it et al.}),{\it Phys. Lett.} {\bf B 39}, 371 (1972).

\bibitem{2} O.B. Abdinov {\it et al.} {\it JINR Rap.Communication},{\bf No 1[75]-96} 51 (1996)

\bibitem{3} BNL E910 Collab. (I. Chemakin {\it et al.})  E-print:
nucl-ex/9902009 (1999)

\bibitem{4} BNL E910 Collab. (Ron Soltz),{\it J. Phys. G: Nucl. Part. Phys.},{\bf 27}
319 (2001)


\bibitem{5} N. Ahababian {\it et al.}, {\it Preprint JINR}, {\bf 1-12114} (1979); N. S.
Angelov {\it et al.}, {\it Preprint JINR}, {\bf 1-12424},(1989);
A.I. Bondarenko {\it et al.}, {\it JINR Communication}, {\bf
P1-98-292} ( 1998); M. K.Suleimanov {\it et al.}, {\it Phys.Rev.}
{\bf C58} 351 (1998).

\bibitem{6} M.C. Abreu {\it et al.}, {\it Phys.Let.} {\bf B450}
456 ( 1999); M.C. Abreu {\it et al.}, {\it Phys.Let.} {\bf B410} 337
(1997); M.C. Abreu {\it et al.}, {\it Phys.Let.} {\bf B410} 327
(1997); NA50 Collab. (M. C. Abreu {\it et al.}) , {\it Phys.Lett.}
{\bf B499} 85 (2001)

\bibitem{percolj} H. Satz, {\it hep-ph/0212046} (2002); Janusz Brzychczyk,{\it nucl-th/0407008} (2004); C. Pajares, {\it hep-ph/0501125} (2005).

\bibitem{7} A.I.Anoshin {\it et al.} {\it Yad.Fiz.} {33} 164(1981)

\bibitem{8} O.B.Abdinov {\it et al.} {\it Preprint JINR},{\bf 1-80-859} (1980)

\bibitem{9} H.H. Heckman {\it et al.} {\it Phys.Rev.} {\bf C17} 1651 (1978)

\bibitem{13} N. Angelov {\it et al.} {\it Preprint JINR}, {\bf P1-11951}
(1978)

\bibitem{14} Harmen J. Warringa.{\it hep-ph/0805.1384} (2008)

\end{thebibliography}
\end{document}